\documentclass[12pt,showpacs,showkeys,preprintnumbers,nofootinbib]{revtex4-1}

\usepackage{amssymb,amsmath,graphics,graphicx}

\begin{document}

\newcommand{\pderiv}[2]{\frac{\partial #1}{\partial #2}}
\newcommand{\deriv}[2]{\frac{d #1}{d #2}}

\title{Impact of contrarians and intransigents in a kinetic model  of opinion dynamics}

\author{Nuno Crokidakis $^{1}$}
\thanks{E-mail: nuno.crokidakis@fis.puc-rio.br}

\author{Victor H. Blanco $^{1}$}

\author{Celia Anteneodo $^{1,2}$}

\affiliation{
$^{1}$Departamento de F\'{\i}sica, PUC-Rio, Rio de Janeiro, Brazil \\
$^{2}$National Institute of Science and Technology for Complex Systems, Rio de Janeiro, Brazil}

\date{\today}

\begin{abstract}
\noindent
In this work we study  opinion formation on a fully-connected population  participating  of a public debate with two distinct choices, where the agents may adopt three different attitudes (favorable  to either one choice or to the other, or  undecided). The interactions between agents occur by pairs and are competitive, with couplings that are either negative with probability $p$ or positive with probability $1-p$. This bimodal probability distribution of couplings produces a behavior similar to the one resulting from the introduction of Galam's contrarians in the population. In addition, we consider that a fraction $d$ of the individuals are intransigent, that is, reluctant to change their opinions. The consequences of the presence of contrarians and intransigents are studied by means of computer simulations. Our results suggest that the presence of  inflexible agents affects the critical behavior of the system, causing either the shift of the critical point or the suppression of the ordering phase transition, depending on the groups of opinions intransigents belong to. We also discuss the relevance of the model for real social systems.

\end{abstract}



\maketitle

\section{Introduction}

Models of opinion formation have been studied by physicists since the 80's and are now part of the new branch of physics called sociophysics \cite{galam_book}. 
This recent research area uses tools and concepts of  statistical physics to describe some aspects of social and political behavior \cite{loreto_rmp}. From the theoretical point of view, opinion models are interesting to physicists because they present order-disorder transitions, scaling and universality, among other typical features of physical systems \cite{loreto_rmp}.

Following the success of the Ising model to capture the essential physics of complex systems, several opinion models have been proposed based on $\pm 1$ (i.e., spin-$1/2$) state variables \cite{loreto_rmp}. The first paper that considered the Ising model to describe a social system was proposed by Galam \cite{galam1982}. The spin-spin coupling of the Ising Hamiltonian represents the agent-agent interaction, whereas the magnetic field represents the effects of propaganda. Moreover, local (or individual) fields are introduced that reflect agent preference toward each orientation (or opinion). Depending on the strength of the local   fields, the system may reach full consensus toward one of the two possible opinions $+1$ or $-1$, or a state in which both opinions coexist. In the last 30 years many other opinion models based on Ising variables have been proposed \cite{galam_book,loreto_rmp}. Among them, we highlight the voter model \cite{liggett,redner}, the majority-rule models \cite{galam_cont,maj_mjo,maj_redner}, the Sznajd model \cite{sznajd} and the CODA (Continuous Opinion and Discrete Actions) model \cite{coda}. Besides the affinity by either one of two distinct opinions or attitudes, one can also consider the possibility that individuals may remain undecided \cite{biswas,meu_celia}. This more realistic situation, that we will consider  here, can be associated  to spin-1 systems, in which the state variables can assume also a null value, besides $\pm 1$.

In order to make the models even more realistic,  other psycho-social ingredients can be taken into account. The so-called \textbf{contrarians} are agents who always have the opposite opinion  to that of the majority of the surrounding agents \cite{galam_cont}. The consideration of such agents affects opinion dynamics, and their impact on opinion formation has been studied in a series of models \cite{biswas,meu_celia,lalama,jorge,galam_hung_elec,kim,kuba}. Another category of agents are the \textbf{intransigents}, whose stubbornness or inflexibility makes them reluctant to change their opinions. This class of agents was firstly introduced in Ref. \cite{moscovici} and  they received later the name inflexible agents or just inflexibles in Ref. \cite{galam_inflex}. After these works, many other papers considered the  effect of inflexibles in opinion dynamics \cite{schneider,jiang,martins,mobilia,galam2011}.

In this work we study a three-state kinetic model of opinion formation, in which the dynamics evolves according to pairwise competitive interactions and where both contrarian and inflexible features are considered. Our results suggest that the presence of inflexible agents affects the critical behavior of the system, 
causing either the shift of the critical point or the suppression of the phase transition, depending on the opinion group  intransigents belong to. 

This work is organized as follows. In Section 2 we present the microscopic rules that define the model. The numerical results are discussed in Section 3, and our conclusions are presented in Section 4.


\section{Model}

Our model is based on kinetic exchange opinion models \cite{biswas,lccc}. 
A population of $N$ agents is defined on a fully-connected graph, i.e.,  each agent can interact  with all others, 
which characterizes a mean-field-like scheme. 
Each individual $i$ ($i=1,2,...,N$) carries one of three possible opinions or attitudes at a given time step $t$, 
represented by $o_{i}(t)=+1,-1$ or $0$. This scenario  mimics any polarized  public debate, 
for example an electoral process with two different candidates A and B, where each agent (or elector) votes for the candidate A (opinion $+1$), 
for the candidate B (opinion $-1$) or remains undecided (opinion $0$). 
In addition, there is a fraction $d$ of agents that are averse to change their opinions, the so-called inflexible agents. 

Each interaction occurs between two given agents $i$ and $j$, such that $j$ will influence $i$. The following rules govern the dynamics:

\begin{enumerate}

\item A pair of agents $(i,j)$ is randomly chosen;

\item If $i$ is an inflexible agent nothing occurs, because he/she cannot be persuaded to change opinion;

\item On the other hand, if $i$ is not an inflexible agent, his/her opinion in the next time step $t+1$ will be updated according to
\begin{equation}\label{eq1}
o_{i}(t+1) = {\rm sgn}\left[ o_{i}(t) + \mu_{ij}\,o_{j}(t)  \right]\,,
\end{equation}
where the sign function is defined such that   ${\rm sgn}(0)=0$  
and the couplings $\{\mu_{ij}\}$ are given by the discrete bimodal probability distribution
\begin{equation}\label{eq2}
F(\mu_{ij}) = p\,\delta(\mu_{ij}+1) + (1-p)\,\delta(\mu_{ij}-1) ~.
\end{equation}

\end{enumerate}

Notice that the above rules impose that for an agent to shift from state $o_{i}=+1$ to $o_{i}=-1$ or vice-versa it must to pass by the intermediate state $o_{i}=0$. The above process is repeated $N$ times, which defines one time step in the simulations. The pairwise couplings may be either negative (with probability $p$) or positive (with probability $1-p$), 
such that $p$ represents the fraction of negative couplings \cite{biswas}. 
In other words, a disorder is introduced in the system, and we will consider that the stochastic random variables $\mu_{ij}$ can be either 
quenched (fixed in time) or annealed (changing with time), as in \cite{biswas,meu_celia}. 
The influence of one individual over another  does not need to be reciprocal (i.e., not necessarily $\mu_{ij}=\mu_{ji}$), 
however, whether interactions are symmetric or not, naturally does not affect the results. The intransigents (a fraction $d$ of the population) are  randomly selected at the beginning of the simulation, 
maintaining that character throughout the dynamics, as considered in the Galam model \cite{galam_inflex}. 

In the absence of intransigents, there is a nonequilibrium order-disorder phase transition at a critical fraction $p_{c}=1/4$ \cite{biswas}. 
For $p<p_{c}$ one of the extreme opinions $+1$ or $-1$ dominates the system, with consensus states occurring only for $p=0$, i.e., 
in the absence of negative interactions. 
On the other hand, for $p\geq p_{c}$ the system is in a  disordered, ``paramagnetic'', phase characterized by the coexistence of the three opinions, 
with the fraction of each opinion being $1/3$. 
Furthermore, it has already been argued \cite{biswas,meu_celia} that negative couplings produce   a similar 
effect to that of the introduction of the Galams' contrarians \cite{galam_cont}, since the main consequence of such negative  couplings is to make that 
interacting agents with the same opinions move to the undecided state (opinion $0$). 
In this sense, our model contains both  contrarian  and inflexible features. 

In the simulations, we have considered two kinds of random couplings $\{\mu_{ij}\}$, quenched and annealed, as well as 
 two kinds of updating schemes,   synchronous (or parallel) and asynchronous (or sequential) updates.
Systems were prepared in fully-disordered initial states, i.e., we started all simulations 
with an equal fraction of each opinion (1/3 for each one). 
In the next section we will present our results.


\section{Results}

We analyze the critical behavior of the system, in analogy to magnetic spin systems, by computing the order parameter  
\begin{equation} \label{eq3}
O = \left\langle \frac{1}{N}\left|\sum_{i=1}^{N} o_{i}\right|\right\rangle ~, 
\end{equation}
where $\langle\, ...\, \rangle$ denotes a disorder or configurational average. 
It is sensitive to the unbalance between extreme opinions. 
Notice that $O$ plays the role of  the ``magnetization per spin'' in  magnetic systems. In addition, we also consider the fluctuations $\chi$ of the order parameter (or ``susceptibility'') 
\begin{equation} \label{eq4}
\chi =  N\,(\langle O^{2}\rangle - \langle O \rangle^{2})   
\end{equation}
and the Binder cumulant $U$, defined as \cite{binder}
\begin{equation} \label{eq5}
U   =   1 - \frac{\langle O^{4}\rangle}{3\,\langle O^{2}\rangle^{2}} \,.
\end{equation}

We analyzed three distinct cases, according to whether  
  the inflexible agents are (i) chosen independently of their initial opinions; 
(ii)    chosen only among the agents with extreme ($\pm 1$) opinions; or (iii)   restricted to a given group of opinion ($o=+1$ or $-1$ or $0$).
In the following subsections, we will present each case separately.


\subsection{Uniformly distributed inflexible agents}

\begin{figure}[t]
\begin{center}
\vspace{3mm}
\includegraphics[width=0.48\textwidth,angle=0]{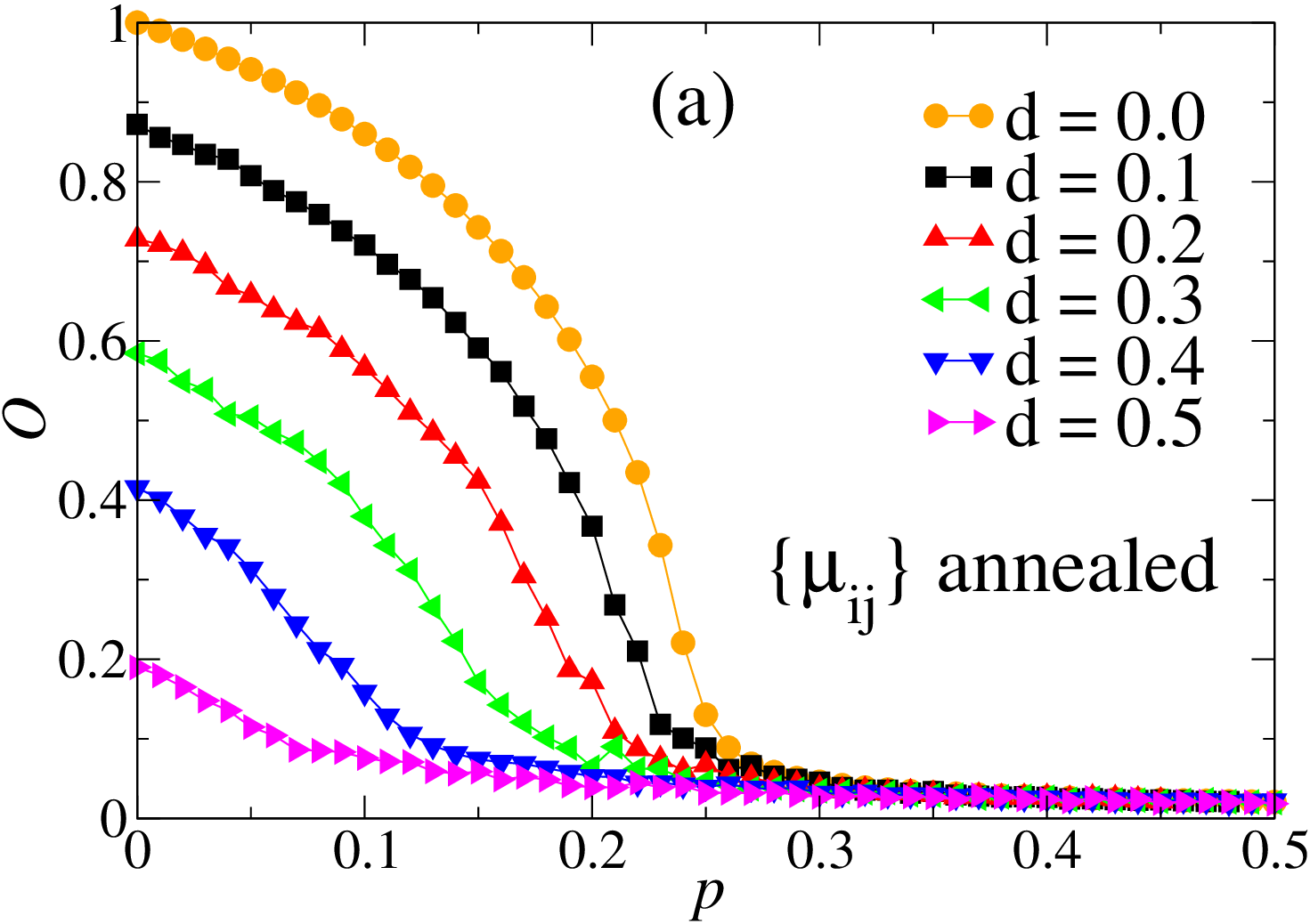}
\hspace{0.3cm}
\includegraphics[width=0.48\textwidth,angle=0]{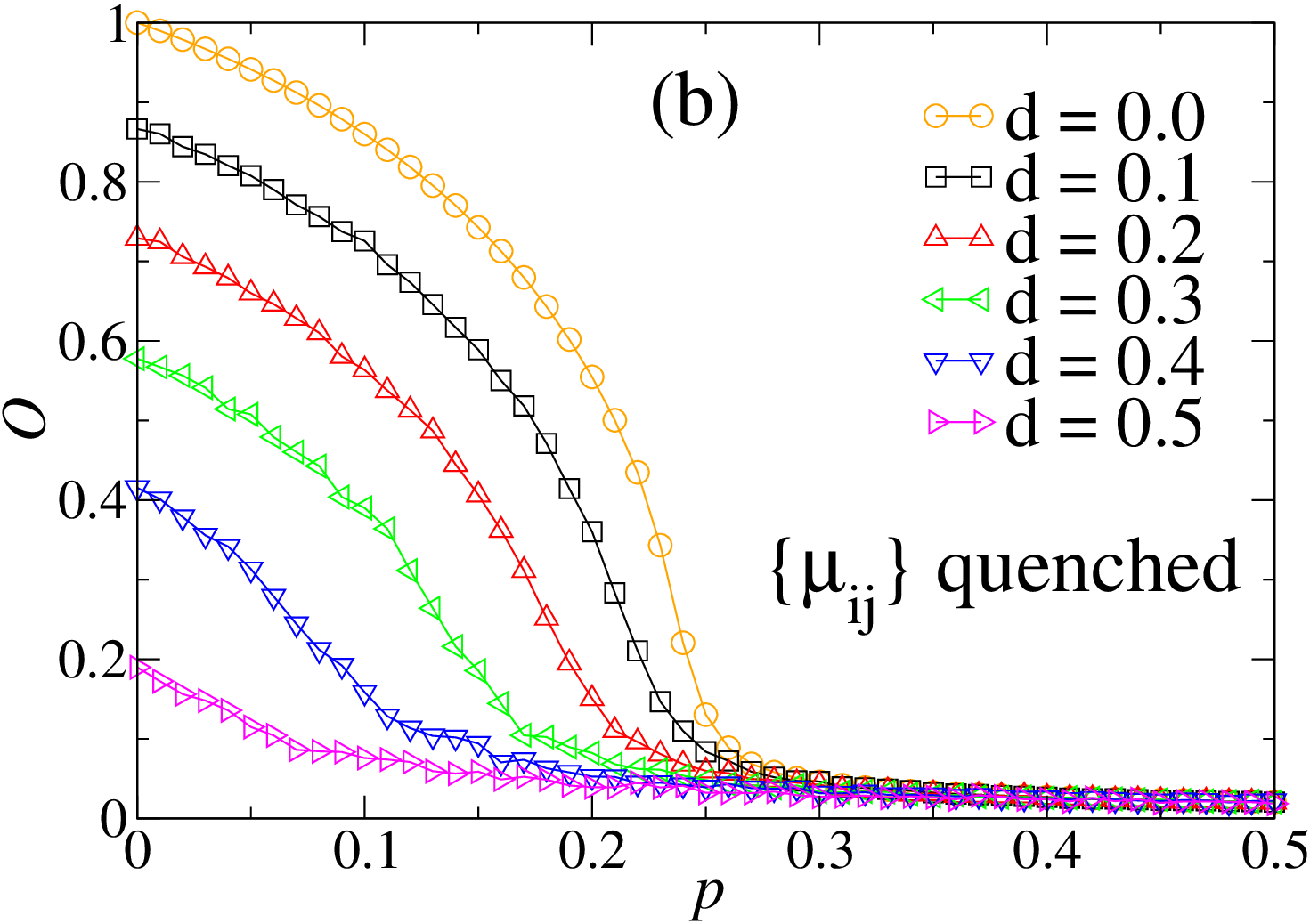}
\end{center}
\caption{(Color online) Order parameter $O$ versus $p$ for typical values of the density $d$ for the case where the inflexibility is independent of the agent opinion. For comparison, we also exhibit the result for the model in the absence of inflexible agents \cite{biswas} ($d=0.0$). The pairwise interactions $\{\mu_{ij}\}$ and the update scheme used are annealed, synchronous (a) and quenched, asynchronous (b), respectively. The population size is $N=1000$ and data are averaged over $100$ simulations.}
\label{fig1}
\end{figure}

In this case, the fraction $d$ of inflexible agents is  randomly selected,   at the beginning of the simulation, 
independently of their opinions. 
In Fig. \ref{fig1} we exhibit the results for the order parameter, Eq. (\ref{eq3}), 
versus the fraction $p$ of negative couplings, for typical values of $d$. 
We display, as examples, the  outcomes for  annealed variables $\{\mu_{ij}\}$ with synchronous updates [Fig. \ref{fig1} (a)], 
and  quenched variables $\{\mu_{ij}\}$ with asynchronous updates [Fig. \ref{fig1} (b)], for a population of $N=1000$ agents. 
Note the strong impact of the change of the parameter $d$ 
on the behavior of the order parameter $O$. Furthermore, given a fixed value of $d$, the curve of $O$ is not affected by the nature of the random variables $\mu_{ij}$ nor by the kind of update scheme used. Consensus states are obtained only in a very specific case: in the absence of both intransigents ($d=0$) and  negative interactions ($p=0$). 
In other words, the maximal value of the order parameter $O$ is smaller than one for all values of $d>0$, independently of $p$. 
In real systems, full consensus, with $O=1$, occurs in particular situations where a government 
exerts a social control, through propaganda or policies that lead to a full acceptance of the \textit{status quo}, 
while collective   states with $O<1$ represent 
more ``democratic'' frequently observed situations \cite{schneider,meu_bjp}. 
Thus, in this sense the inclusion of inflexible agents makes the model more realistic.

For sufficiently large $d$ the system is always found in a disordered (paramagnetic) phase, but for small values of $d$ 
the system orders   at specific points that depend on $d$.  
In order to locate the critical points $p_{c}(d)$ numerically, we have performed simulations for different population sizes $N$. 
Thus, the transition points $p_{c}(d)$ are estimated, for each value of $d$, from the crossing of the Binder cumulant 
curves for the different sizes \cite{binder}. In addition, a finite-size scaling (FSS) analysis was performed, 
in order to obtain an estimate of the critical exponents $\beta$, $\gamma$ and $\nu$, by means of the usual FSS equations
\begin{eqnarray} \label{eq6}
O(d,N) & \sim & N^{-\beta/\nu} \\  \label{eq7}
\chi(d,N) & \sim & N^{\gamma/\nu} \\   \label{eq8}
U(d,N) & \sim & {\rm constant} \\   \label{eq9}
p_{c}(d,N) - p_{c}(d) & \sim & N^{-1/\nu} ~,
\end{eqnarray}
that are valid in the vicinity of the transition. 

\begin{figure}[t]
\begin{center}
\includegraphics[width=0.47\textwidth,angle=0]{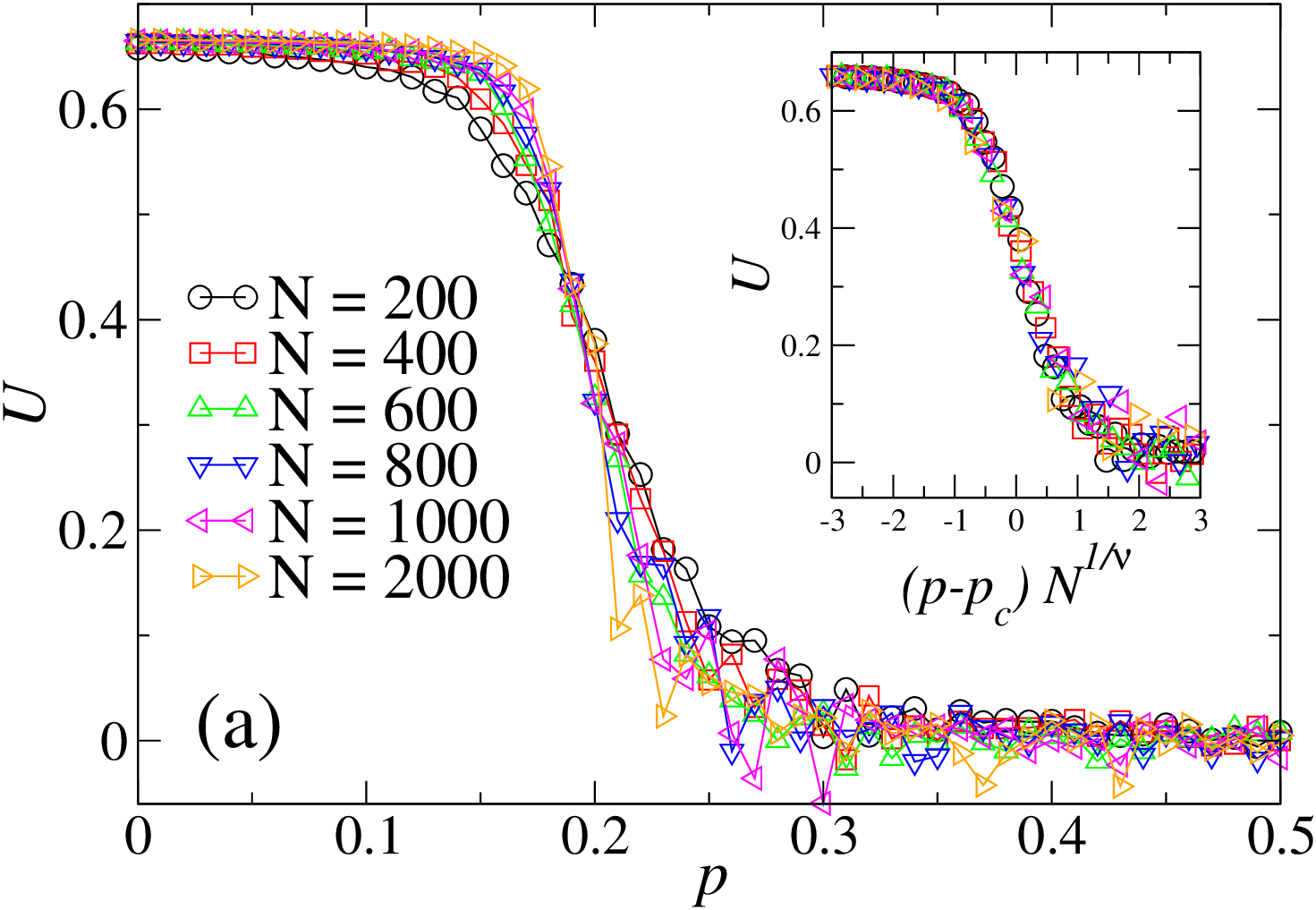}
\hspace{0.5cm}
\includegraphics[width=0.47\textwidth,angle=0]{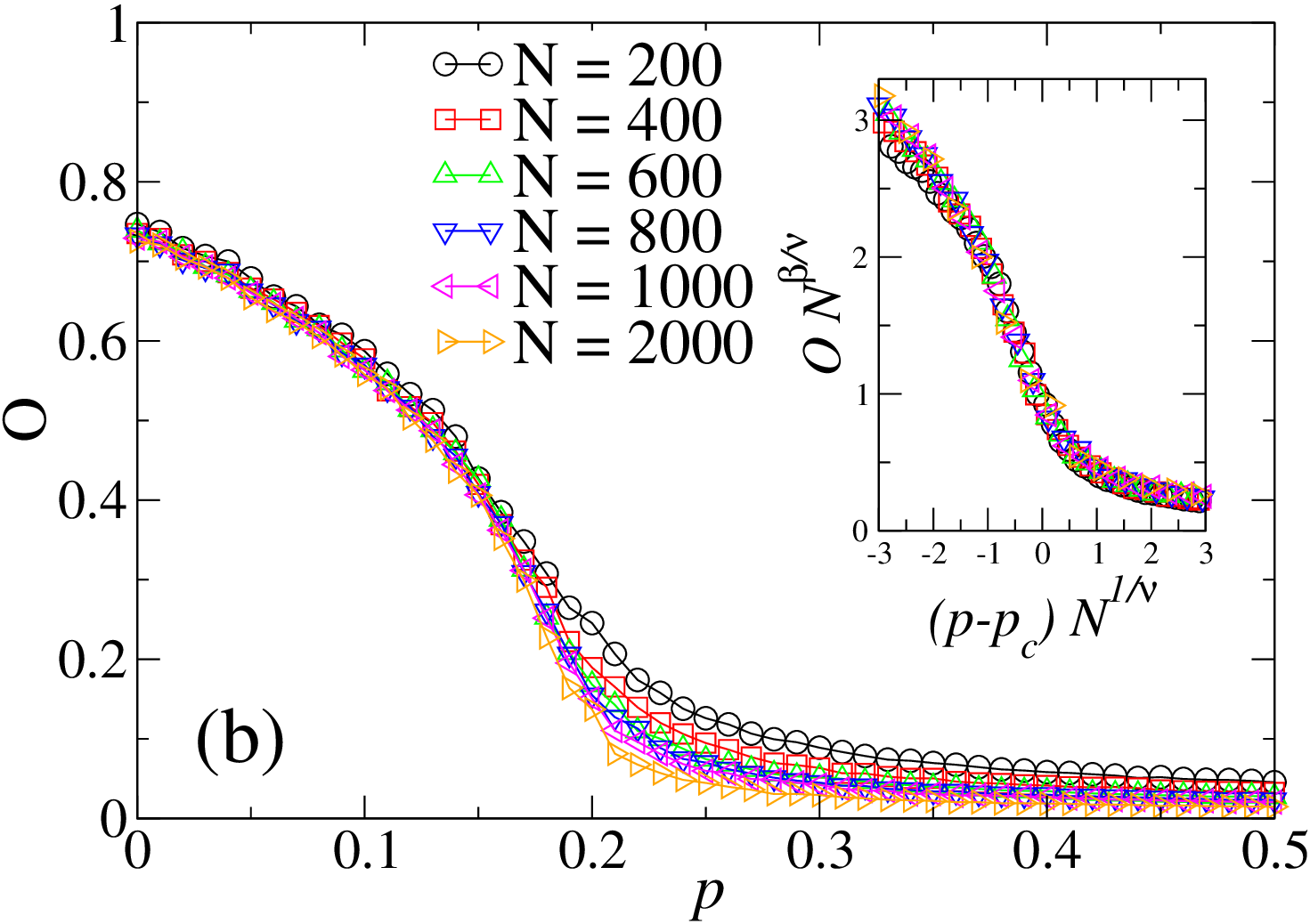}
\\
\vspace{0.9cm}
\includegraphics[width=0.47\textwidth,angle=0]{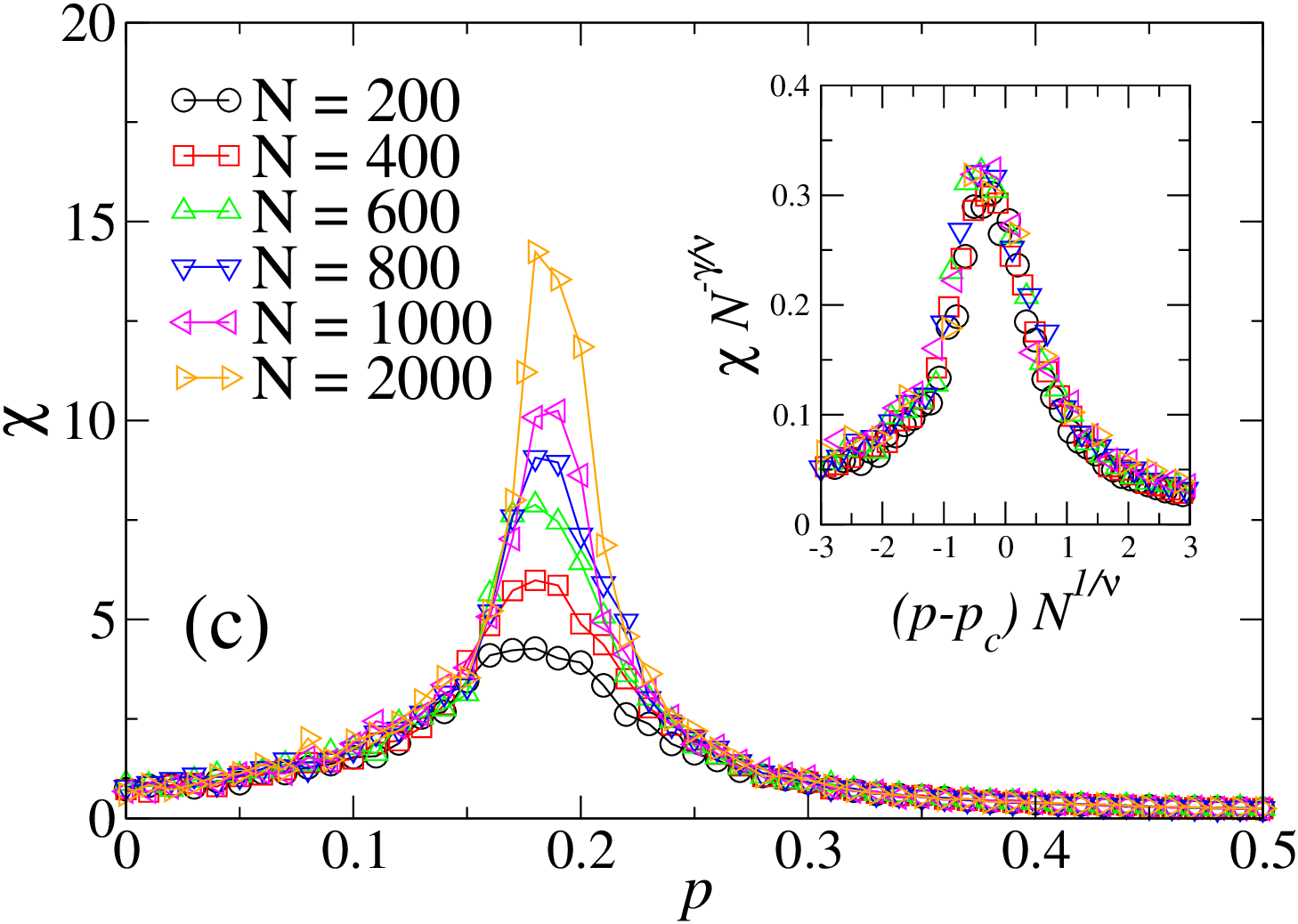}
\end{center}
\caption{(Color online) Binder cumulant (a), order parameter (b) and susceptibility (c) for the case where the inflexibility is independent of the agent opinion, for $d=0.2$ and different population sizes $N$ (main plots). The corresponding scaling plots are shown in the respective insets. Data are for quenched random variables $\{\mu_{ij}\}$ and asynchronous update scheme. The best data collapse was obtained for $p_c=0.196$, $\beta=0.5$, $\gamma=1.0$ and $1/\nu=0.5$.}
\label{fig2}
\end{figure}

As an illustration, we exhibit in Fig. \ref{fig2} 
the behavior of the quantities of interest as well as the scaling plots for $d=0.2$, quenched random couplings 
and asynchronous updates. Our estimates for the critical exponents coincide with those for the original model ($d=0$), 
i.e., we obtained $\beta\approx 0.5$, $\gamma\approx 1.0$ and $1/\nu\approx 0.5$. 
These exponents are robust: they are the same, within error bars estimated from the FSS analysis, for all values of $d$, 
independently of the update scheme considered and of the kind of random couplings $\{\mu_{ij}\}$ (quenched or annealed).

\begin{figure}[t]
\begin{center}
\vspace{0.7cm}
\includegraphics[width=0.48\textwidth,angle=0]{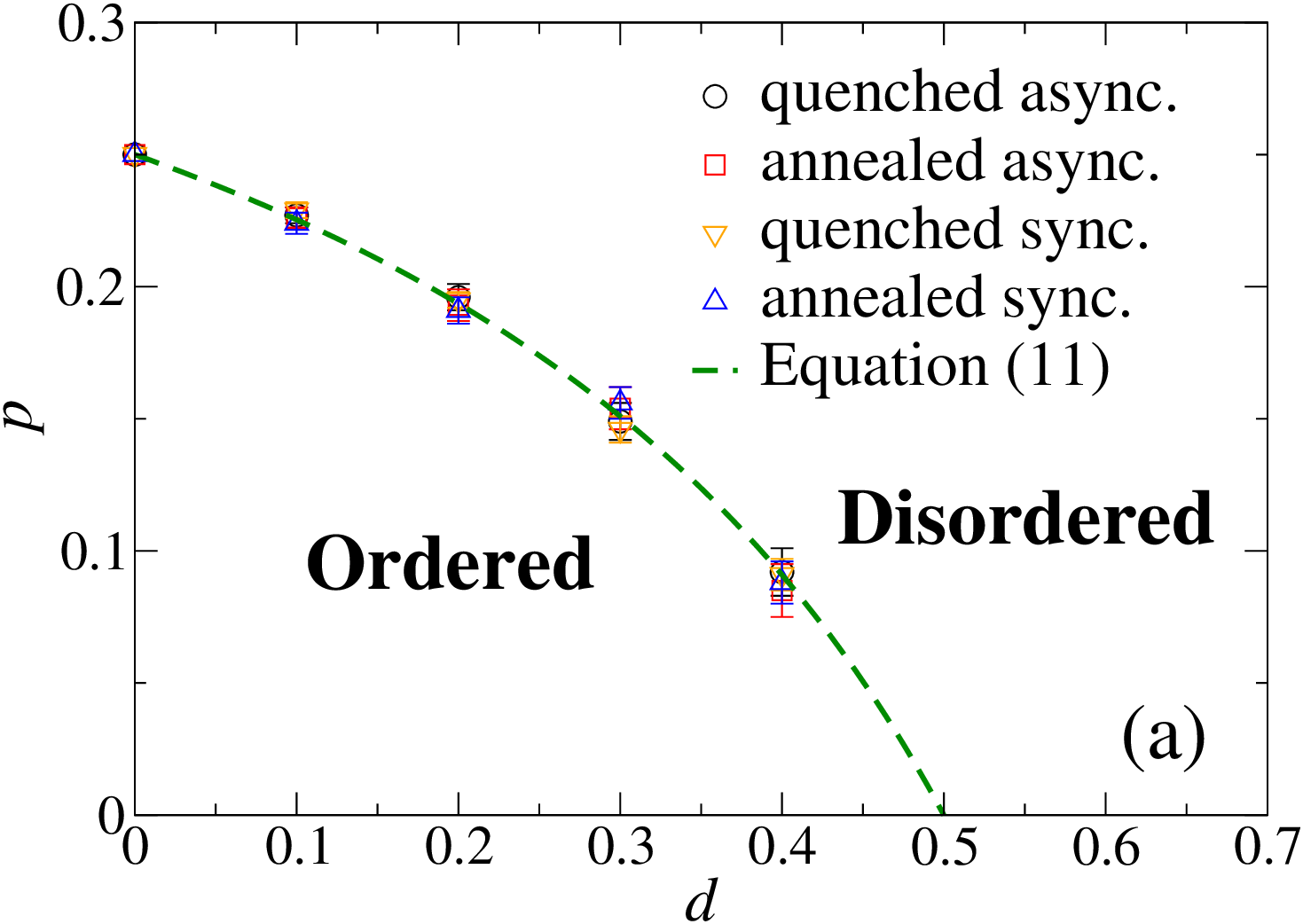}
\hspace{0.3cm}
\includegraphics[width=0.48\textwidth,angle=0]{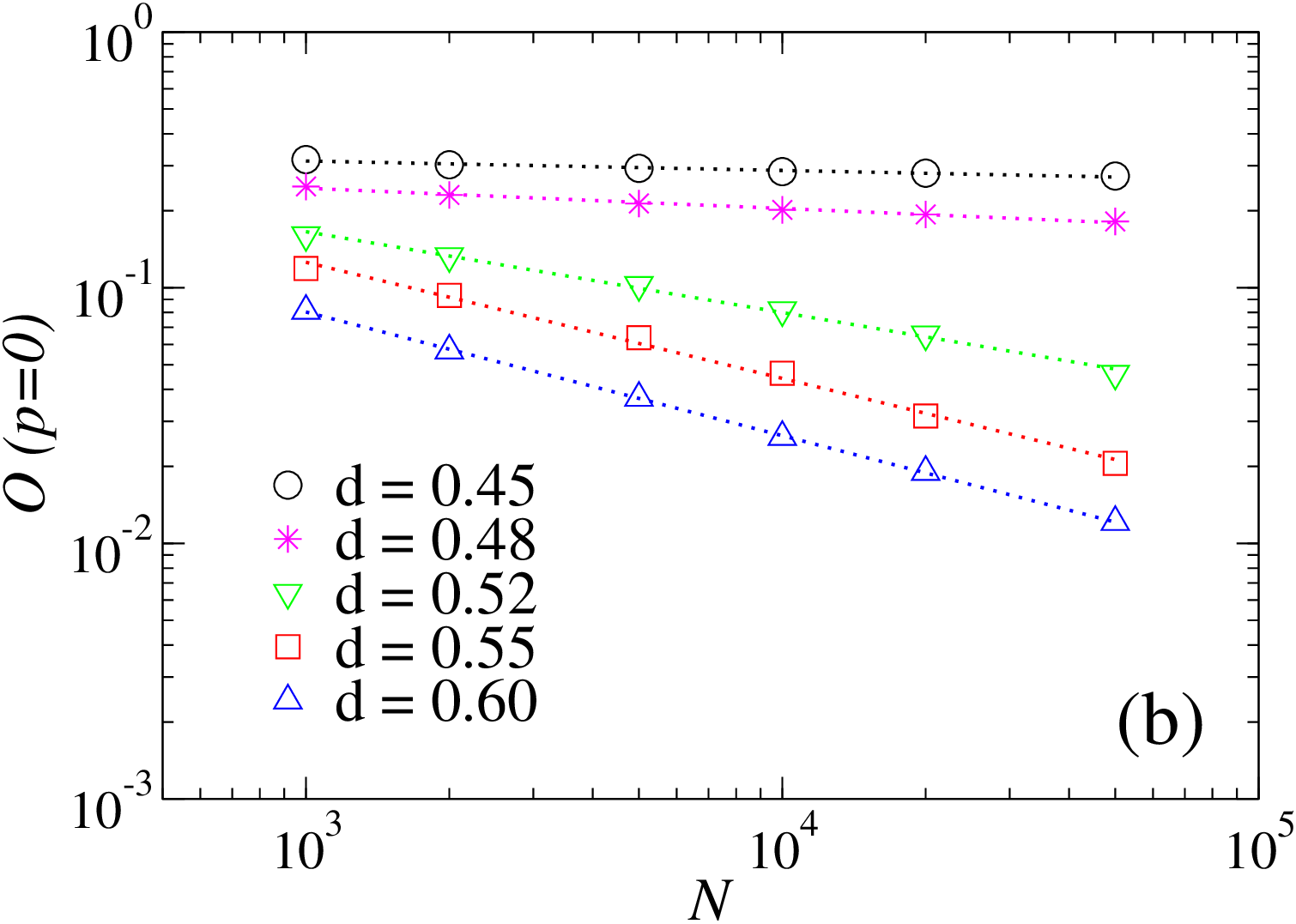}
\end{center}
\caption{(Color online) (a) Phase diagram of the model in the plane $p$ versus $d$ for the case where the inflexibility is independent of the agent opinion. We exhibit results for quenched and annealed couplings $\{\mu_{ij}\}$ and synchronous and asynchronous updates. The dashed line is a qualitative description of the phase boundary given by Eq. (\ref{eq11}), as discussed in the text. (b) Maximal values of the order parameter as a function of the population size $N$, in the log-log scale, for values of $d$ near the critical density $d_{c}\approx 0.5$. }
\label{fig3}
\end{figure}

Taking into account the FSS analysis for typical values of $d$, we exhibit in Fig. \ref{fig3} (a) the phase diagram 
of the model in the plane $p$ versus $d$. The symbols are the numerical estimates for the critical points $p_{c}(d)$. 
Based on the analytical results of the annealed version of a similar model \cite{meu_celia}, where the critical points 
are given by a ratio of two first-order polynomials, we propose the following qualitative form for the critical frontier,
\begin{equation}\label{eq10}
p_{c}(d) = \frac{x\,d+1}{y\,d+4}  ~,
\end{equation} 
where $x$ and $y$ are real numbers, and we have taken into account the analytical result of the model in 
the absence of inflexible agents, $p_{c}(d=0)=1/4$ \cite{biswas}. Fitting the numerical values of $p_{c}(d)$ with Eq. (\ref{eq10}), we obtained 
\begin{equation}\label{eq11}
p_{c}(d) = \frac{2\,d-1}{4.5\,d-4}  ~.
\end{equation} 

\noindent
Eq. (\ref{eq11}) is plotted in Fig. \ref{fig3} (a) together with the numerical results. One can see that the curve describes qualitatively well the phase boundary between the ordered and the disordered phase, and the simulation data agrees within error bars with Eq. (\ref{eq11}). Based on Eq. (\ref{eq11}) one can estimate the critical density $d_{c}$ above which the system cannot order. This critical value is $d_{c}\approx 0.5$, and above it the three opinions $+1$, $-1$ and $0$ coexist in the population (1/3 in average for each one), which is a characteristic of the disordered phase of this kind of model \cite{biswas,meu_celia}. 
In order to test the validity of Eq. (\ref{eq11}) and the estimated value of $d_{c}$, we simulated the system for $p=0$ and different population sizes $N$, and we measured the order parameter $O(p=0)$. One can see in Fig. \ref{fig3} (b) that for $d<0.5$ the values of $O(p=0)$ remain almost constant for increasing sizes. Nonetheless, for $d>0.5$ the values of $O(p=0)$ decreases as a power-law of $N$, which indicates that we have $O(p=0)=0$ in the thermodynamic limit. Thus, these results suggest that the system will be in a disordered paramagnetic phase for $d>0.5$. As discussed above, the critical exponents are the same for all values of $d$, indicating a universality on the order-disorder frontier. Thus, for sufficiently large values of the fraction of inflexible agents, the order-disorder transition is eliminated.


\subsection{Inflexible agents restricted to the extreme opinions}

We consider a variant of the model considered in the previous section.
 Instead of selecting as intransigent agents a fraction $d$ of the population totally at random, one can restrict the inflexibility to agents that initially have one of the extreme opinions $\pm 1$. In other words,  with probability $d$ an agent  is set as inflexible  only if this agent has initial opinion either $o=+1$ or $o=-1$. This is also a realistic case, since in some countries there are intransigents supporting  two extreme opinions, while the remaining individuals are undecided or intend to nullify their votes. In this case, these ``neutral'' agents can be persuaded by the decided individuals and adopt one of the extreme opinions (e.g., left or right candidate).

\begin{figure}[t]
\begin{center}
\includegraphics[width=0.48\textwidth,angle=0]{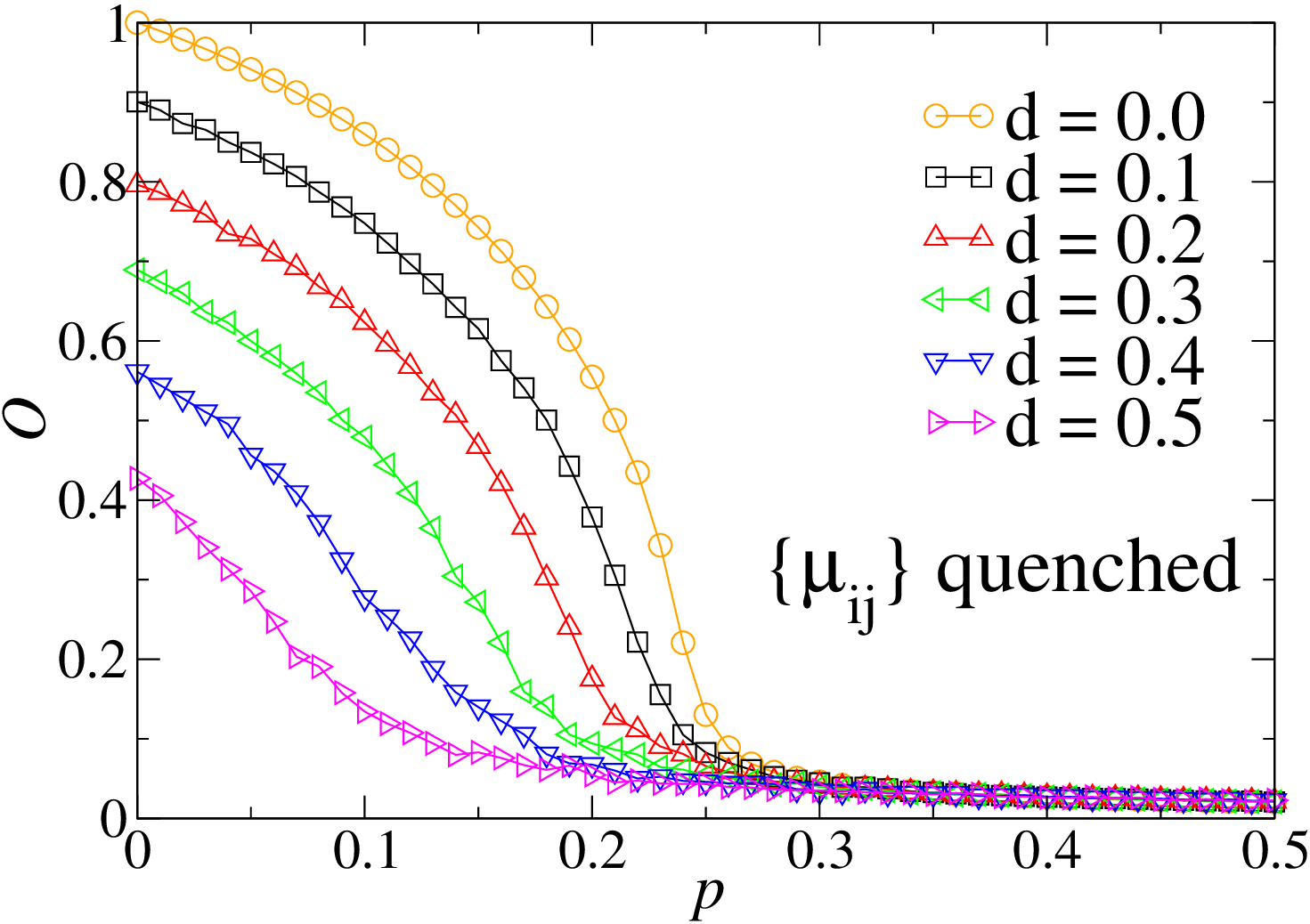}
\end{center}
\caption{(Color online) Order parameter $O$ versus $p$ for typical values of $d$ for the case where the inflexibility is associated with agents carrying the extreme opinions ($o=\pm 1$). For comparison, we also exhibit the result for the model in the absence of inflexible agents \cite{biswas} ($d=0.0$). The pairwise interactions $\{\mu_{ij}\}$ and the update scheme used are quenched and asynchronous, respectively. The population size is $N=1000$ and data are averaged over $100$ simulations.}
\label{fig4}
\end{figure}

As in the previous case, we first studied the behavior of the order parameter $O$ as a function of the fraction $p$ of negative interactions, for typical values of the density $d$. In Fig. \ref{fig4} we exhibit a representative example, for the case where the random couplings $\{\mu_{ij}\}$ are quenched variables and the states are updated in a asynchronous way. One can see that the qualitative behavior is similar to the one presented in the last subsection (III.A), i.e., we observe order-disorder transitions at different values $p_{c}$ that depend on $d$. However, there are qualitative differences, and a comparison with Fig. \ref{fig1} (b) shows that the increase of $d$ affects the order parameter less in the case where the intransigents are restricted to agents with the extreme opinions $\pm 1$. As a consequence, the values of $p_{c}(d)$ are different from those observed in the previous subsection. This fact can be easily understood. The agents with extreme opinions (be they inflexible or not) can provoke a change of the $0$ opinions to $+1$ or $-1$, which does not happen if the undecided agents are allowed to be  intransigents. This fact favors the ordering in the system, and so the magnetization per spin in the actual case presents greater values than in the case where the inflexibility is not restricted and can be a characteristic of the individuals independently of their opinions. Once again, the nature of $\{\mu_{ij}\}$ and the update scheme used do not affect the results.

\begin{figure}[t]
\begin{center}
\vspace{0.7cm}
\includegraphics[width=0.48\textwidth,angle=0]{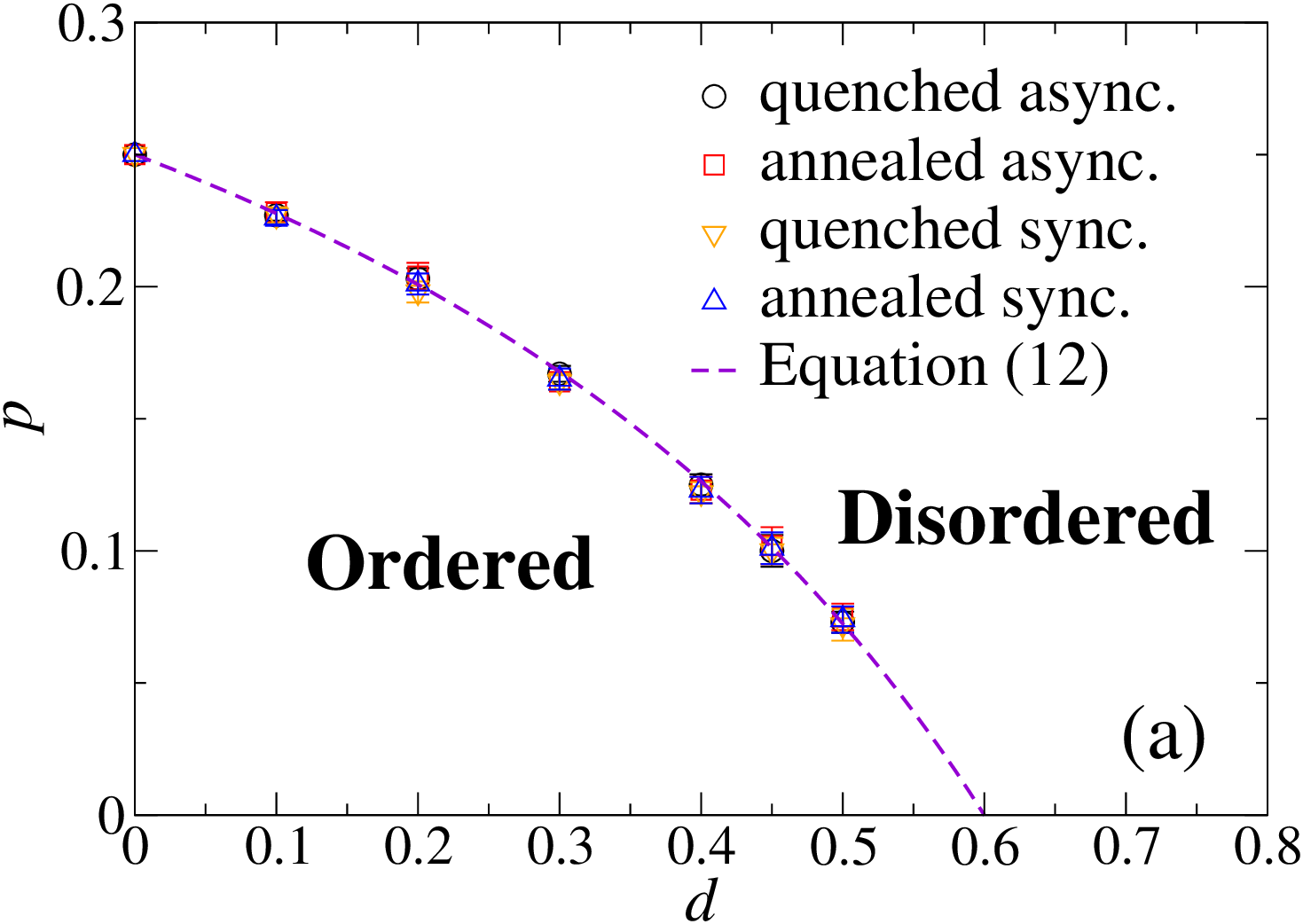}
\hspace{0.3cm}
\includegraphics[width=0.48\textwidth,angle=0]{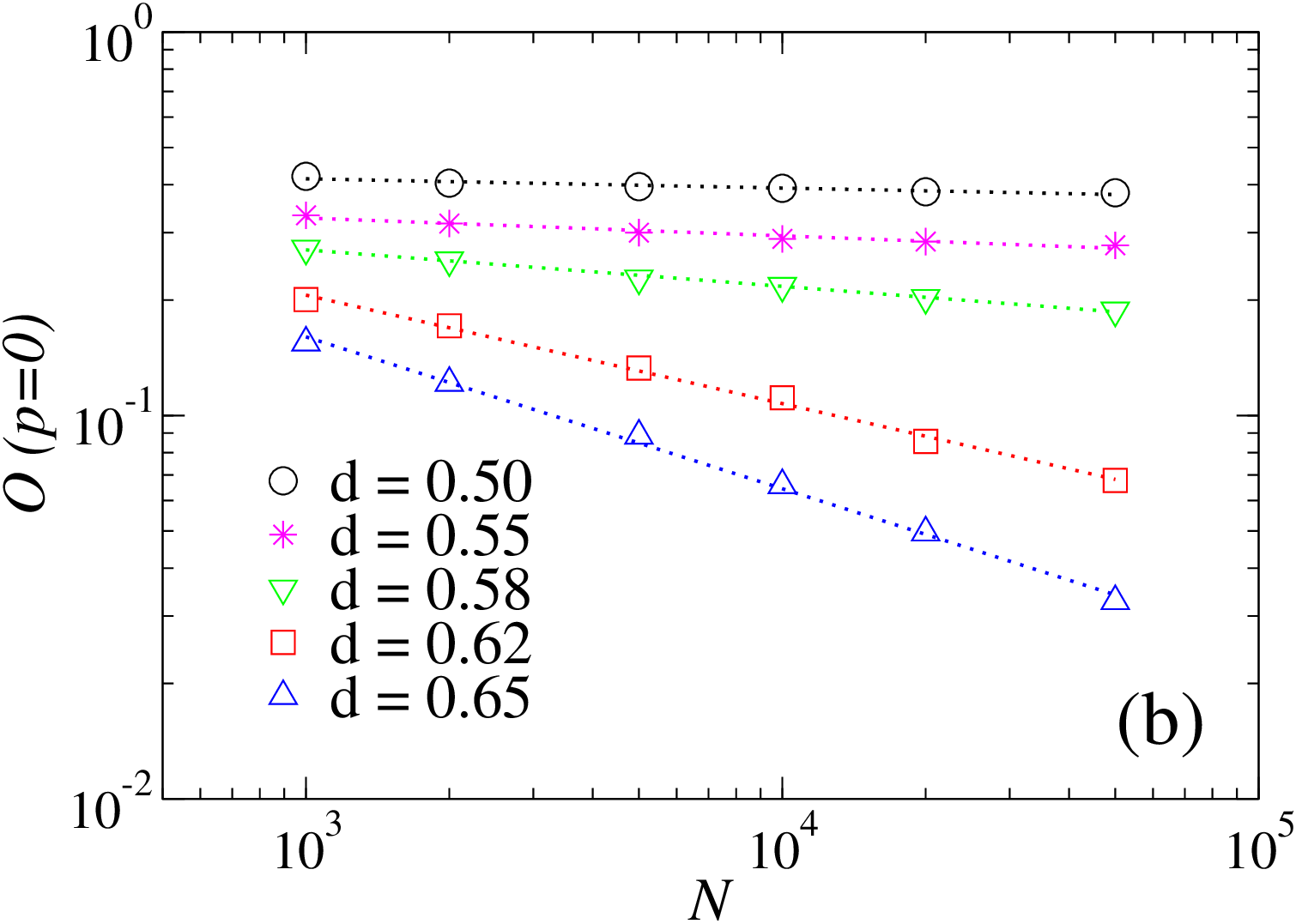}
\end{center}
\caption{(Color online) (a) Phase diagram of the model in the plane $p$ versus $d$ for the case where the inflexibility is associated with the extreme opinions ($o=\pm 1$). We exhibit results for quenched and annealed couplings $\{\mu_{ij}\}$ and synchronous and asynchronous updates. The dashed line is a qualitative description of the phase boundary given by Eq. (\ref{eq12}), as discussed in the text. (b) Maximal values of the order parameter as a function of the population size $N$, in the log-log scale, for values of $d$ near the critical density $d_{c}\approx 0.6$.}
\label{fig5}
\end{figure}

As in the previous section, we performed a FSS analysis in order to obtain the critical points $p_{c}(d)$ and the critical exponents $\beta$, $\gamma$ and $\nu$. In Fig. \ref{fig5} (a) we exhibit the phase diagram of the model in the plane $p$ versus $d$. The symbols are the numerical estimates for the critical points $p_{c}(d)$. Again, we propose the qualitative form of Eq. (\ref{eq10}) for the order-disorder frontier. Fitting data, we obtained 
\begin{equation}\label{eq12}
p_{c}(d) = \frac{1.67\,d-1}{3.4\,d-4}  ~.  
\end{equation} 
\noindent
In other words, we have a similar frontier than in the previous case, but with different parameters. Eq. (\ref{eq12}) is plotted in Fig. \ref{fig5} (a) together with the numerical results. One can see that the curve describes qualitatively well the phase boundary between the ordered and the disordered phase, and the simulation data agree within error bars with Eq. (\ref{eq12}). 

Based on Eq. (\ref{eq12}) one can estimate the critical density $d_{c}$ 
above which the system cannot order. This critical value is $d_{c}\approx 0.6$. 
Thus, for sufficiently large values of the fraction of inflexible agents, the order-disorder transition is eliminated. Notice that the critical density in this case ($d_{c}\approx 0.6$) is greater than the critical density of the previous case ($d_{c}\approx 0.5$), where the intransigent agents may be chosen independently of their opinions. The origin of this difference is again related to the agents with $o=0$ opinions, as discussed in the beginning of this section. In fact, as the presence of non-inflexible agents with $o=0$ opinions favors the ordering in the system, as discussed above, the critical fraction $p_{c}$ becomes larger in the present case than in the case where the inflexibility is not restricted and can be a characteristic of a given individual independently of his opinion. As in the previous section, we performed simulations for the system with $p=0.0$ and different population sizes $N$ in order to test the validity of the estimated value of $d_{c}$. One can see in Fig. \ref{fig5} (b) that for $d<0.6$ the values of $O(p=0)$ remain almost constant for increasing sizes. Nonetheless, for $d>0.6$ the values of $O(p=0)$ decreases as a power-law of $N$, which indicates that we have $O(p=0)=0$ in the thermodynamic limit. Thus, these results reinforce the idea that the system will be in a disordered paramagnetic phase for $d>0.6$. It is important to say that we obtained the usual exponents $\beta\approx 0.5$, $\gamma\approx 1.0$ and $1/\nu\approx 0.5$ for all values of $d$ considered in Fig. \ref{fig5} (a), which indicates that the universality class of the model is not affected when we consider inflexible agents only among the individuals with the extreme opinions $o=\pm 1$.


\subsection{Inflexible agents restricted to a given opinion}

\begin{figure}[t]
\begin{center}
\vspace{0.4cm}
\includegraphics[width=0.48\textwidth,angle=0]{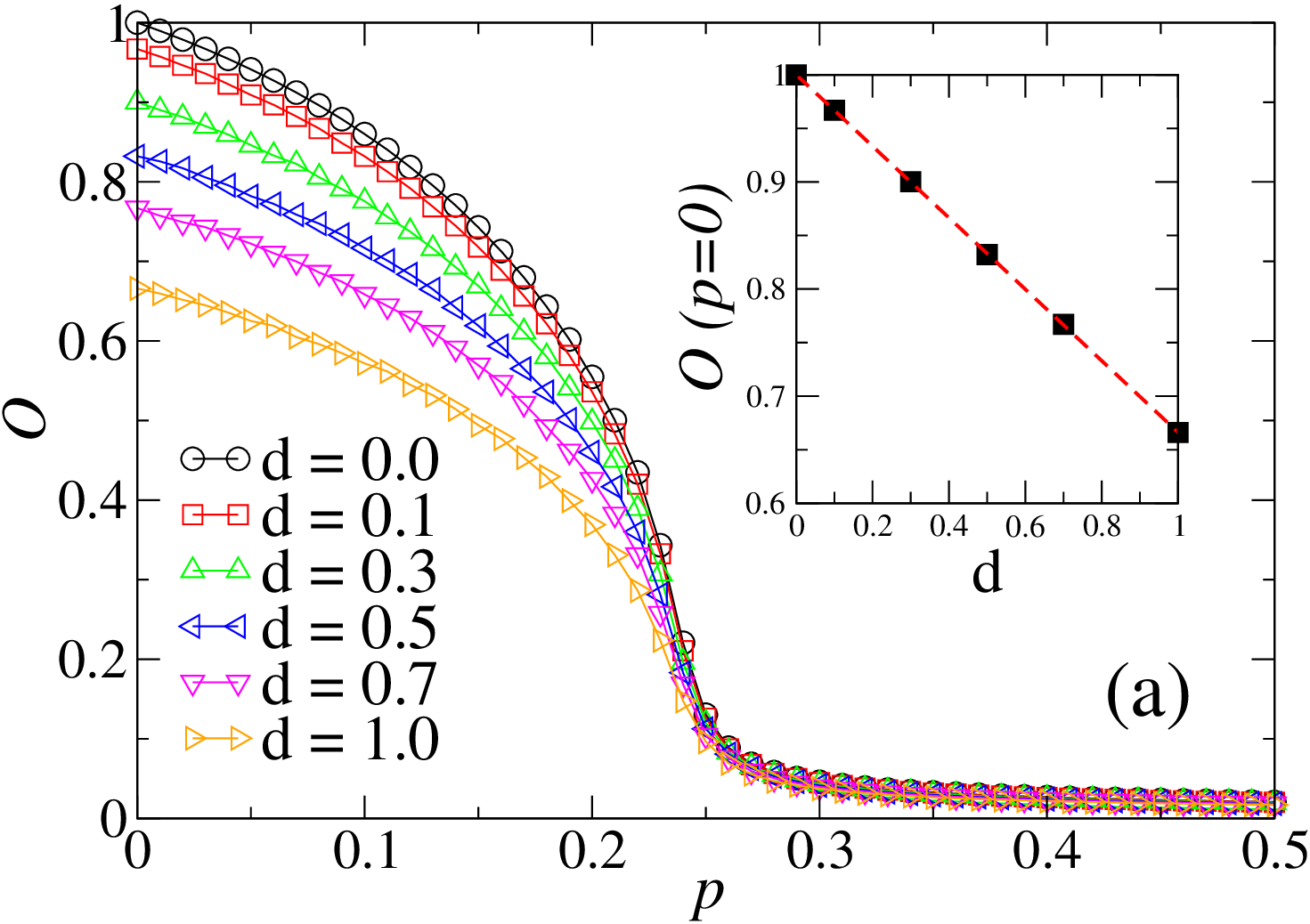}
\hspace{0.3cm}
\includegraphics[width=0.48\textwidth,angle=0]{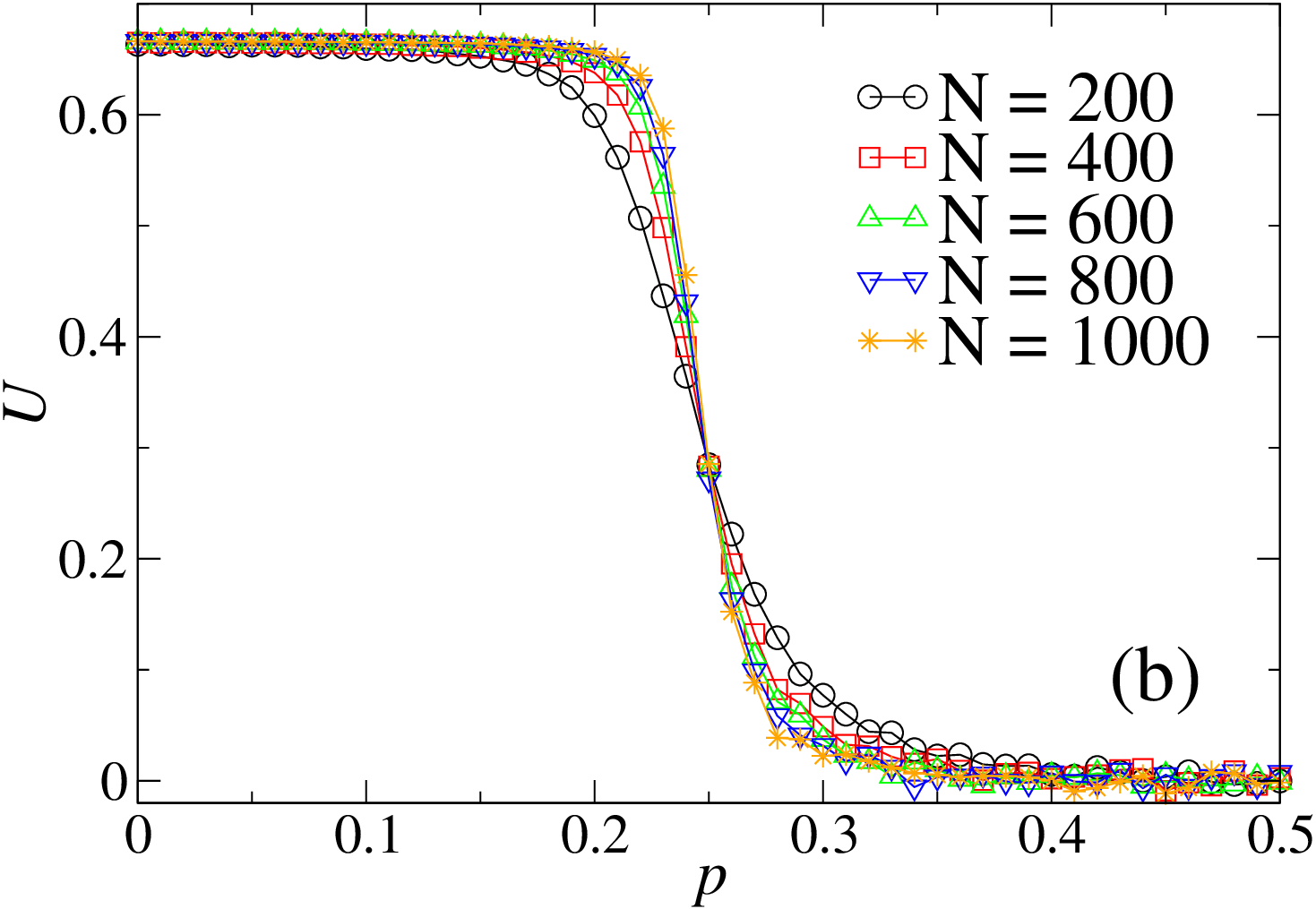}
\end{center}
\caption{(Color online) Results for the case where the intransigents are chosen among the agents with opinion $o=0$. (a) Order parameter as a function of $p$ for typical values of $d$, for population size $N=1000$. The inset shows the maximal value of the order parameter for a given value of $d$, that occurs for $p=0$. Fitting data, we obtained $O(p=0)=1 - d/3$. (b) Binder cumulant for $d=1.0$ and different sizes $N$, showing a crossing at $p_{c}\approx 0.25$. In both cases the interactions $\{\mu_{ij}\}$ are quenched random variables, and we have considered asynchronous updates.}
\label{fig6}
\end{figure}

Finally, we also study another variant of the model considered in Section III.A. Instead of selecting the $d\,N$ inflexible agents  at random, one can restrict the selection to a given opinion group. In other words, with probability $d$ an agent  becomes inflexible, but now only if this agent has initial opinion either $o=+1$ or $o=-1$ or $o=0$. This can  also be a realistic situation, since in some countries there is only a certain point of view (or opinion) that is shared by an intransigent group. 

We can first consider the case where the intransigents are chosen among the agents  with opinion $o=0$. We exhibit in Fig. \ref{fig6} (a) the results for the order parameter as a function of $p$ for typical values of $d$. One can see that the maximal value of the order parameter (for $p=0$) decreases for increasing values of $d$. This result is expected, since the initial condition is fully disordered (1/3 of each opinion), and a fraction $d$ of the agents with opinion $o=0$ are selected as intransigents at $t=0$. In this case, the maximum   of the order parameter should be $O_{{\rm max}}=O(p=0)=1-d/3$, which is confirmed by the simulations (see the inset of Fig. \ref{fig6} (a)). Although the values of the order parameter for $p < p_{c}$ are different for distinct values of $d$, the order-disorder transition occurs at the same point. An example is given in Fig. \ref{fig6} (b), where we exhibit the Binder cumulant as a function of $p$ for the maximum   of the density of intransigents, $d=1.0$. One can observe a crossing of the curves at $p_{c}\approx 0.25$. We also performed a FSS analysis (not shown), which confirms the same exponents observed in the previous sections, i.e., we have $\beta\approx 0.5$, $\gamma\approx 1.0$ and $1/\nu\approx 0.5$. These results are independent of $d$, which indicates that the universality class of the model is not affected when we consider inflexible agents only among the individuals with opinion $o=0$.

\begin{figure}[t]
\begin{center}
\vspace{0.4cm}
\includegraphics[width=0.48\textwidth,angle=0]{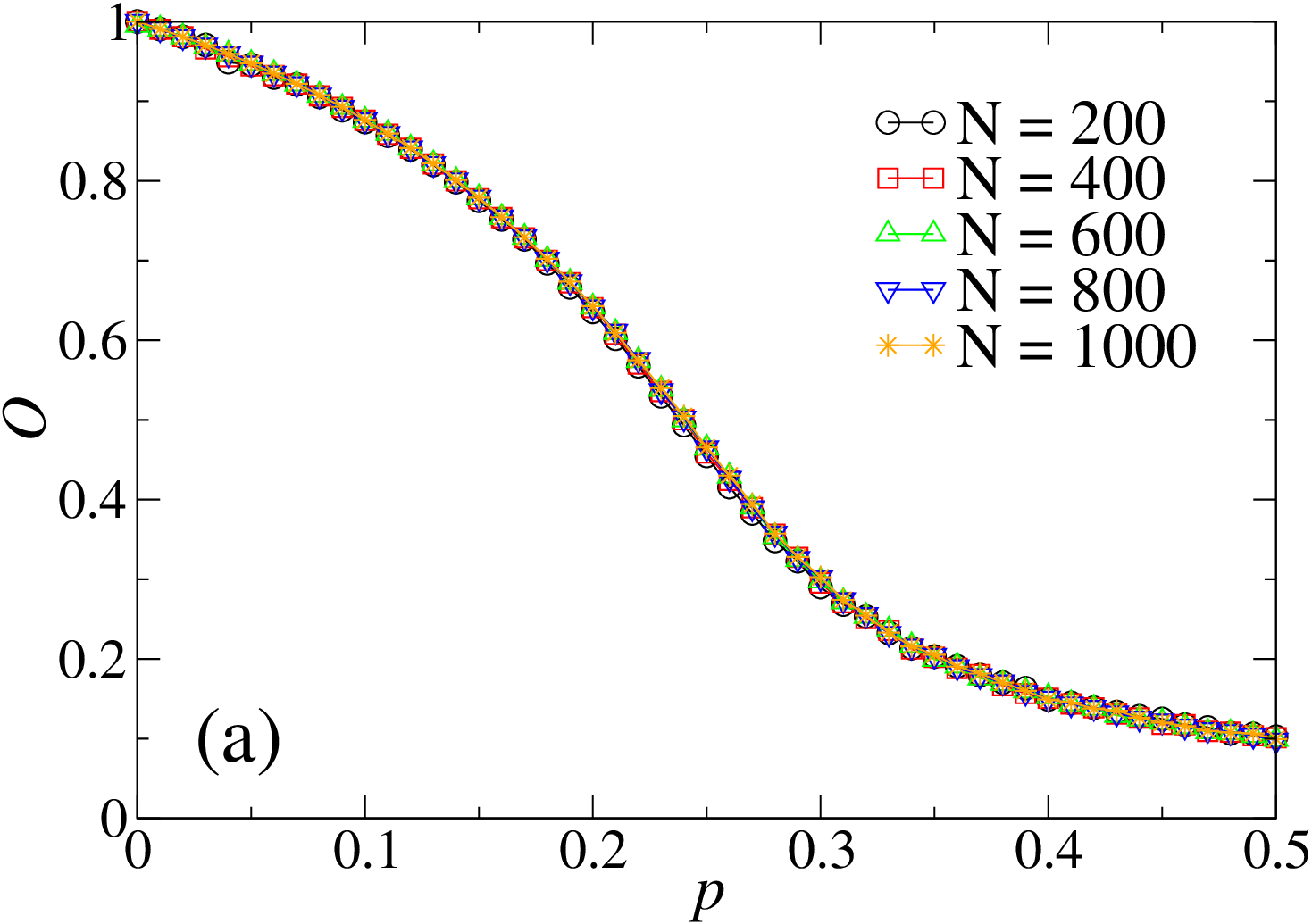}
\hspace{0.3cm}
\includegraphics[width=0.48\textwidth,angle=0]{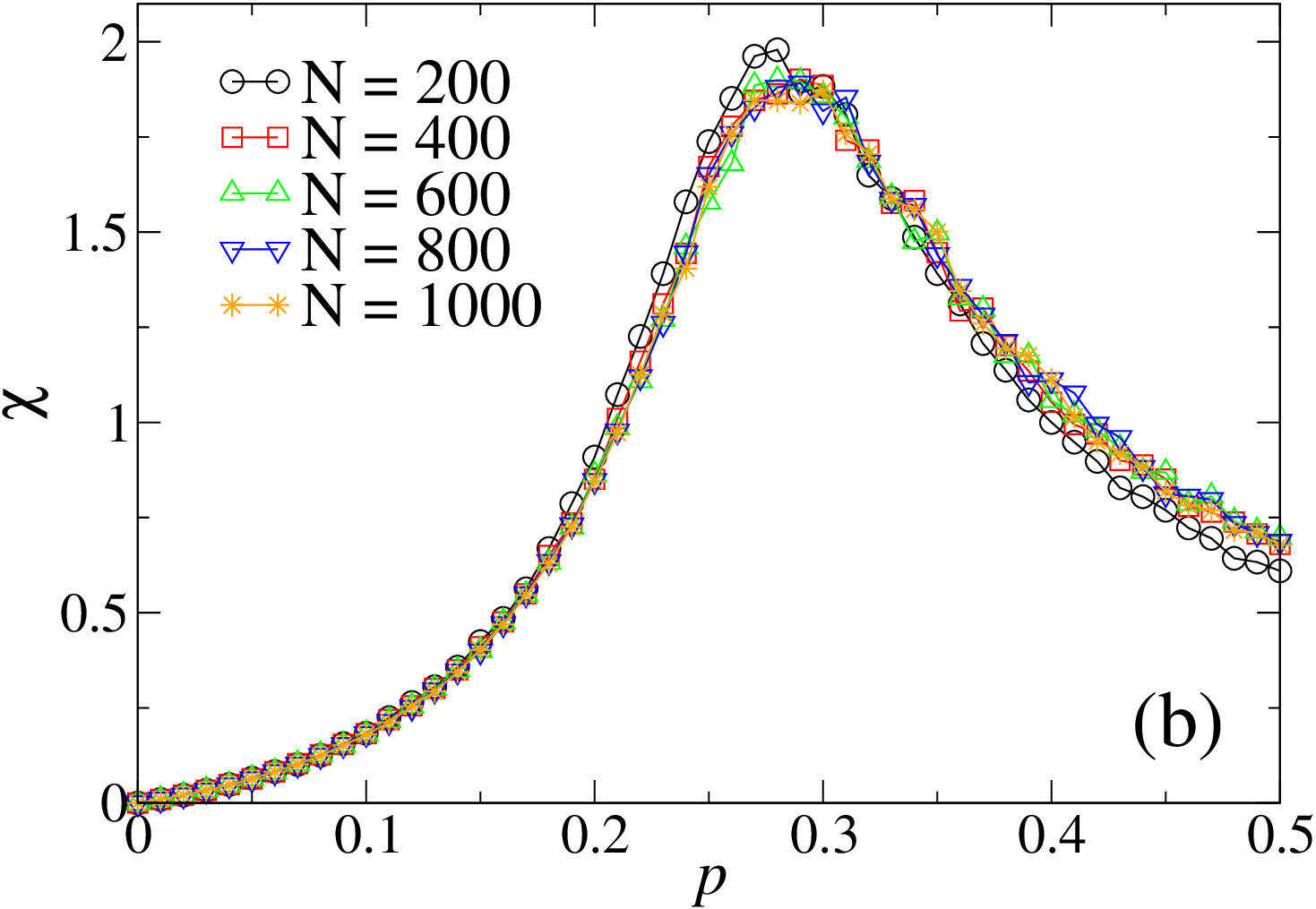}
\end{center}
\caption{(Color online) Results for the case where the inflexible agents are chosen among the agents with opinion $o=+1$, for $d=0.3$. (a) Order parameter and (b) susceptibility as functions of $p$ for different population sizes $N$. Notice that there is no dependence of the results on the system size. In both cases the interactions $\{\mu_{ij}\}$ are quenched random variables, and we have considered asynchronous updates.}
\label{fig7}
\end{figure}

In the case where the intransigents are restricted to agents with opinion $o=+1$ 
\footnote{The behavior of the model in the case where they are restricted to agents with opinion $o=-1$ is identical to the case $o=+1$.}, the results are different from the previous case. We have observed that the order parameter decays with increasing values of $p$, as usual, but the lower values of $O$ are not so small as usual (see Fig. \ref{fig7} (a)). 
\begin{figure}[t]
\begin{center}
\vspace{0.6cm}
\includegraphics[width=0.48\textwidth,angle=0]{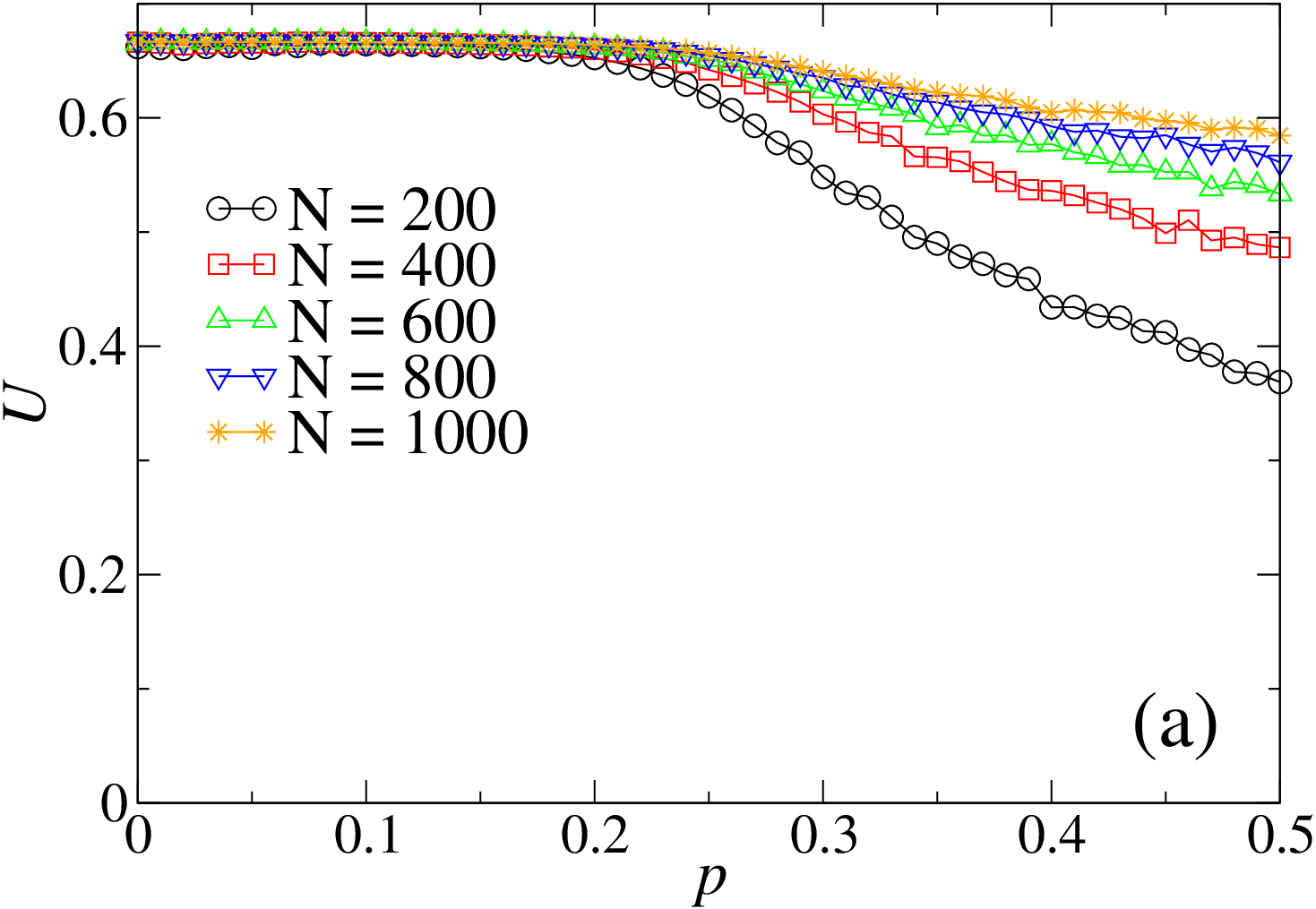}
\hspace{0.3cm}
\includegraphics[width=0.48\textwidth,angle=0]{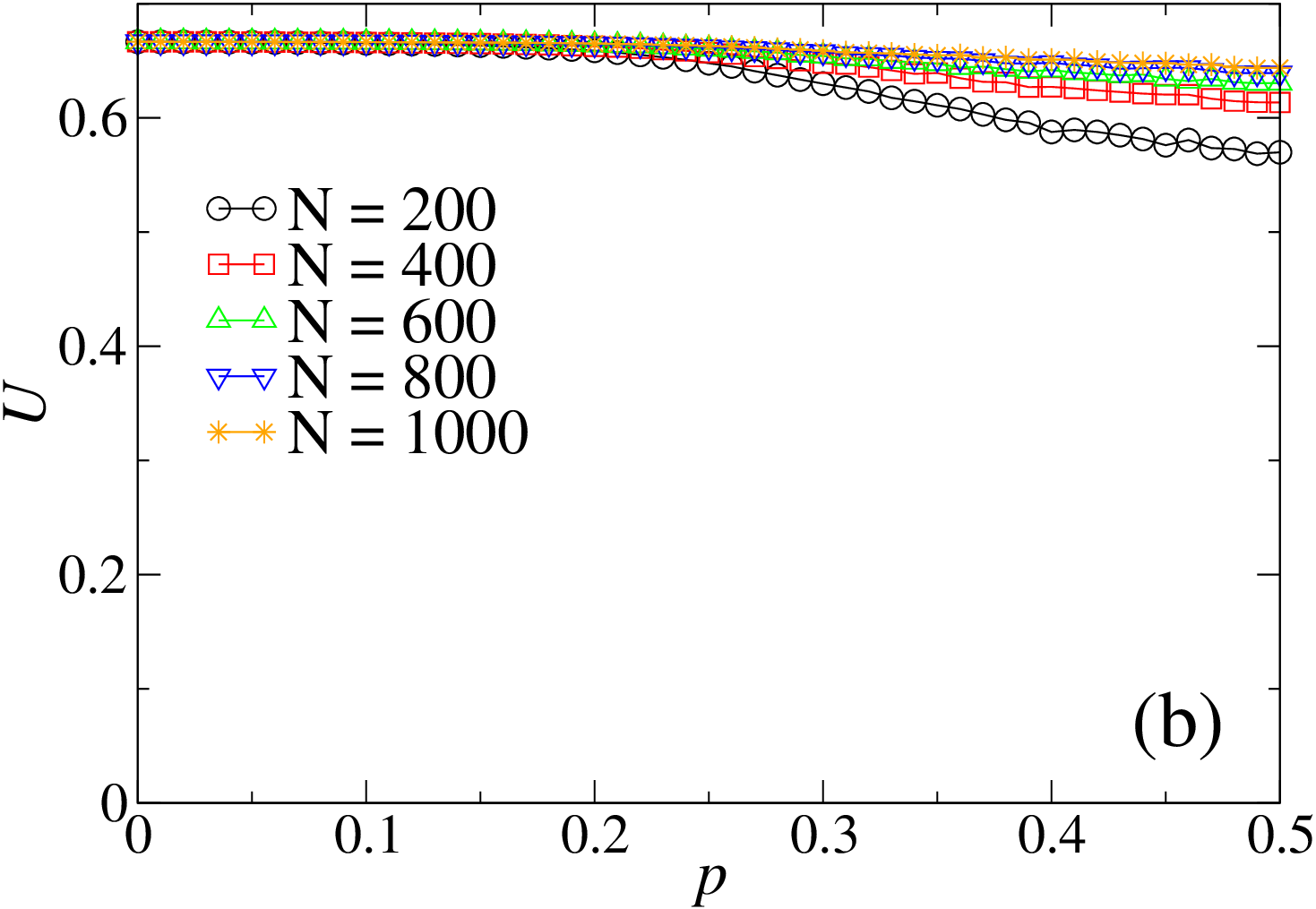}
\end{center}
\caption{(Color online) Binder cumulant for the case where the intransigents are chosen among the agents with opinion $o=+1$, for different population sizes $N$. (a) $d=0.3$ and (b) $d=0.6$. Notice the absence of the crossing of the curves. In both cases the interactions $\{\mu_{ij}\}$ are quenched random variables, and we have considered asynchronous updates.}
\label{fig8}
\end{figure}

In addition, the order parameter curves, as well as the susceptibility ones, do not depend on the system size (see Fig. \ref{fig7}), as usually occurs in phase transitions \cite{landau_book,dickman_book}. These results suggest that there is no phase transition when we consider inflexible agents only among agents with one of the extreme opinions, $o=+1$ or $o=-1$. To confirm this picture, we plot in Fig. \ref{fig8} the Binder cumulant for two different values of $d$, namely $d=0.3$ (a) and $d=0.6$ (b), and different sizes $N$. We can observe that in both cases the Binder cumulant curves do not cross, indicating that there is no phase transition \cite{binder}. Notice also from Fig. \ref{fig8} that the absence of 
the phase transition is more pronounced for higher values of $d$. Thus, one can conclude that there is a crossover in the population, i.e., the order parameter decreases when we rise the fraction of negative interactions $p$, but there are no divergences associated with this crossover, suggesting the absence of a phase transition.

Summarizing this section, our results show that when we consider the inflexible agents distributed only among the agents carrying a given opinion, the critical behavior is identical to that of the model in the absence of intransigents ($d=0$) \cite{biswas}, when the considered opinion is $o=0$, i.e., we have a transition at $p_{c}=1/4$. On the other hand, if the inflexibility is related to the extreme opinions $o=+1$ or $o=-1$, the phase transition is suppressed.


\section{Final remarks}   

In this work, we have studied a discrete-state opinion model where each agent carries one of three possible opinions or attitudes, represented by variables $+1$, $-1$ and $0$. The microscopic rules define that agents with the extreme opinions $\pm 1$ should pass by the intermediate (undecided) state $0$ before adopting the opposite extreme opinion. We have considered a population of $N$ agents in the mean-field limit, where each individual can interact with all others. The competitive interactions, ruled by negative (with probability $p$) and positive (with probability $1-p$) couplings, produce an effect similar to Galam's contrarians. Moreover, a fraction $d$ of the population  is constituted by intransigents, averse to change their opinions. In this sense, our model takes into account both contrarians and intransigents in the process of opinion formation.

The subset of inflexible agents (a fraction $d$ of the population) is randomly selected at the beginning of the simulation,  keeping the inflexible character throughout the dynamics. This is a realistic social feature.  Indeed, intransigent individuals usually do not change their attitude with time. 

We have analyzed cases where the inflexiblity is not restricted to a given opinion group, as well as cases where inflexibility is associated to the extreme ($\pm 1$) opinions or to a given group supporting one of the three possible attitudes. Moreover, we have also considered that the agents' states (opinions) are updated by means of either sequential (asynchronous) or parallel (synchronous) schemes. 

In the first formulation of the model, the inflexible agents are chosen independently of their opinions. By analyzing the quantities of interest (magnetization per spin, susceptibility and Binder cumulant), we have found that the system exhibits continuous nonequilibrium phase transitions between an ordered phase and a disordered one. The transition points depend on the density $d$ of intransigents, similarly to what happens in other models \cite{galam_inflex,jiang}. The simulations show that the values of $p_{c}(d)$ decrease for increasing values of $d$, hence the disordered phase broadens with increasing $d$. Numerical outcomes suggest that there is a critical density $d_{c}\approx 0.5$ above which the system cannot order, i.e., the system is in a fully-disordered (paramagnetic) state, for all values of $p$. The critical exponents on the order-disorder frontier are the same, $\beta=1/2$, $\gamma=1$ and $\nu=2$, independently of $d$, which means  a universality in the model. These results are not affected by the update scheme used (synchronous or asynchronous) nor by the nature of the random couplings (quenched or annealed).

In the second formulation of the model, the inflexible agents are chosen only among the agents with (initially) extreme opinions. In this case, the model behavior is qualitatively similar to the previous one. However,  the critical density in this case is greater, $d_{c}\approx 0.6$. Thus, the ordered phase is larger when the agents with $o=0$ opinions are free to interact, which is the main fact responsible for the observed differences. However, the critical behavior of the model is robust with respect to the selection of the inflexible agents. In fact, the critical exponents on the order-disorder frontier are the same as in the previous case, $\beta=1/2$, $\gamma=1$ and $\nu=2$, independently of $d$. This confirms the universal behavior of that phase transition. Again, this result is not affected by the update scheme used (synchronous or asynchronous) nor by the nature of the random interactions (quenched or annealed).

We have also considered the case where the inflexible agents are chosen among the individuals with a given initial opinion. For the case where this opinion is $o=0$, the critical behavior of the system is not affected by the presence 
of the intransigent agents, i.e., the phase transition occurs at $p_{c}=1/4$ for all values of $d$. On the other hand, when the intransigents are chosen among the agents with opinion $o=+1$ (or alternatively, $o=-1$), the phase transition does not occur anymore. This conclusion was supported by the behavior of the quantities of interest. In fact, the order parameter and the susceptibility curves do not depend on the system size, and the Binder cumulant curves for different population sizes do not cross. All these features suggest the absence of the order-disorder transition \cite{landau_book,dickman_book}.

Notice that in the cases where the phase transition occurs, the critical exponents are always the same, $\beta=1/2$, $\gamma=1$ and $\nu=2$. This is an expected result, since we are dealing with a mean-field formulation of the model, where each agent can interact with all others. Observe that the values of $\beta$ and $\gamma$ are the same as the mean-field exponents of the Ising model, but the exponent $\nu$ presents a different value. As discussed in Ref. \cite{biswas}, interpreting $\nu$ as $\nu^{'}\,D$ where $D$ is the effective dimension in this long-ranged model and considering this effective dimension as $D=4$, then the value of the effective correlation length exponent becomes $\nu^{'}=1/2$, that coincides with the mean-field value.

Our results also show that the particular nature of the random couplings $\mu$, as well as of the update scheme,  does not affect the results. This may seem an obvious result, however, in other variants of the kinetic exchange opinion model, we observed that the results can be affected by numerical considerations like the fluctuation or not of the pairwise interactions $\mu$ (annealed and quenched versions, respectively), or the synchronous or asynchronous update scheme, as  recently shown in Ref. \cite{meu_celia}.

Despite the simplicity of our model, it can be relevant for the description of real social systems. In our model, $o=1$ represents a favorable opinion and $o=-1$ an unfavorable one, while $o=0$ means indecision. The order parameter considered corresponds to the overall rating and an ordered state means there is a clearcut decision made. A disordered state means the absence of a decision. Thus, the contrarian effect, quantified by the parameter $p$, induces a disordered phase for sufficiently large $p$. In addition, the inclusion of inflexible agents, quantified by the parameter $d$, makes this effect more pronounced, since the critical points $p_{c}$ decrease for increasing values of $d$. Thus, the presence of such two effects, contrarians and intransigents, favors the disordered state, indicating that in the presence of extremists it is more difficult to reach a final decision in a public debate, which is a realistic feature of the model. In addition, the results show that the consensus states are never obtained when inflexible agents are present. This is also realistic in elections or public debates in general. In fact, the occurrence of consensus states with the order parameter $O=1$ occurs in very particular situations, whereas the states with partial order ($O<1$) are more common \cite{schneider,meu_bjp}. 

We hope that theoretical opinion models considering realistic individuals like contrarians and intransigents may also guide proper new experiments (such as inquiries or surveys) to be conducted for improving the construction of agent-based models, as well as for the validation of such models.


\section*{Acknowledgments}

The authors acknowledge financial support from the Brazilian scientific funding entities CAPES, FAPERJ and CNPq.

\end{document}